\documentclass[
preprint]{ptptex}

\newcommand{\tr}{\mathop{\rm tr}\nolimits}

\newcommand{\Det}{\mathop{\rm Det}\nolimits}
\newcommand{\Pf}{\mathop{\rm Pf}\nolimits}
\newcommand{\SU}{\mathop{\rm SU}}
\newcommand{\SO}{\mathop{\rm SO}}

\newcommand{\rmd}{{\rm d}}

\newcommand\fverb{\setbox\pippobox=\hbox\bgroup\verb}
\newcommand\fverbdo{\egroup\medskip\noindent%
                        \fbox{\unhbox\pippobox}\ }
\newcommand\fverbit{\egroup\item[\fbox{\unhbox\pippobox}]}
\newbox\pippobox

\preprintnumber[3cm]{
NIIG-DP-05-3\\RIKEN-TH-57\\{\tt hep-th/0511207}}

\title{
Zero-dimensional Analogue of the Global Gauge Anomaly%
}

\author{
Hiroto \textsc{So}$^{1,}$\footnote{E-mail: so@muse.sc.niigata-u.ac.jp}
and Hiroshi \textsc{Suzuki}$^{2,}$\footnote{E-mail: hsuzuki@riken.jp}
}

\inst{
$^1$Department of Physics, Niigata University, Ikarashi 2-8050,\\
Niigata, 950-2181, Japan\\
$^2$Theoretical Physics Laboratory, RIKEN, Wako 2-1, Saitama 351-0198, Japan
}

\abst{
A zero-dimensional analogue of Witten's global gauge anomaly is considered.
For example, a zero-dimensional reduction of the two-dimensional $\SO(2N)$
Yang-Mills theory with a single Majorana-Weyl fermion in the fundamental
representation suffers from this anomaly. Another example is a
zero-dimensional reduction of two- and three-dimensional $\SU(2N_c)$
Yang-Mills theories which couple to a single Majorana fermion in the adjoint
representation. In this case, any expectation value is either indeterminate
or infinite.}

\begin{document}

\maketitle

In the present note, we consider integrals of the form
\begin{equation}
   \int\rmd\mu(A)\,\mathcal{O}(A)\int\rmd\chi\,e^{\chi^TA\chi}
   =\int\rmd\mu(A)\,\mathcal{O}(A)\Pf(A),
\label{one}
\end{equation}
where $A$ is a $2N\times 2N$ anti-symmetric matrix, $\chi$ is a
Grassmann-odd $2N$-column vector and $\mathcal{O}(A)$ is a function of~$A$.

Suppose that there exists a certain matrix~$g\in\mathop{\rm {}O}(2N)$ such
that
\begin{equation}
   \rmd\mu(g^TAg)=\rmd\mu(A),\qquad
   \mathcal{O}(g^TAg)=\mathcal{O}(A).
\label{two}
\end{equation}
Suppose further that $\det g=-1$. Then we have
\begin{equation}
   \Pf(g^TAg)=(\det g)\Pf(A)=-\Pf(A),
\label{three}
\end{equation}
where the first equality follows from the definition of the pfaffian (or
by considering the jacobian associated with the change of the integration
variable~$\chi\to g^{-1}\chi$), and hence the integral~(\ref{one})
identically vanishes. The expectation value of the
function~$\mathcal{O}(A)$, defined by
\begin{equation}
   \langle\mathcal{O}\rangle=
   {\displaystyle
   \int\rmd\mu(A)\,\mathcal{O}(A)\int\rmd\chi\,e^{\chi^TA\chi}
   \over
   \displaystyle
   \int\rmd\mu(A)\int\rmd\chi\,e^{\chi^TA\chi}},
\label{four}
\end{equation}
is then indeterminate, being $0/0$.

The integral~(\ref{one}) can be regarded as a functional integral in a
zero-dimensional gauge field theory in which a ``Majorana fermion'' $\chi$
is coupled to a ``gauge potential'' $A$. The Grassmann variable~$\chi$ is
regarded as a Majorana spinor, because the conjugate
variable~$\overline\chi$ does not appear in the integral. Interestingly, this
simple integral represents an analogue of Witten's global gauge
anomaly\cite{Witten:fp}. Firstly, for the global gauge anomaly to exist,
there must exist an element of the gauge transformation that cannot be
continuously deformed into the identity. This requires that $\pi_d(G)$ be
non-trivial in $d$~dimensions, where $G$ is the gauge group. In zero
dimensions, this would correspond to $\pi_0(G)\neq0$, which requires that
$G$ have disconnected components. The group~$\mathop{\rm {}O}(2N)$ in fact
has two components, one with $\det g=+1$ and another with $\det g=-1$. For
elements in a component connected to the identity, we have the invariance of
the Majorana pfaffian expressed by $\Pf(g^TAg)=\Pf(A)$ (which can be
regarded as an analogue of the absence of the local gauge anomaly for
Majorana fermions), while for elements in the component with~$\det g=-1$, we
have the non-invariance expressed by $\Pf(g^TAg)=-\Pf(A)$. One may also
consider a spectral flow of eigenvalues of~$A$ along a one-parameter family
$A_t=(1-t)A+tg^TAg$. The pfaffian $\Pf(A)$, which is the square root of
$\det A$, changes sign along the one-parameter family~$A_t$. Secondly, in
$8k$~dimensions, a single Majorana fermion cannot have a Lorentz invariant
mass term, and there is no obvious way to apply a gauge invariant
Pauli-Villars regularization. In fact, the Majorana fermion in
$8k$~dimensions is equivalent to the Weyl fermion in a real representation,
and the latter may suffer from the global gauge anomaly\cite{Holman:ef}.
(Pseudo-real representations, by contrast, are free from the global gauge
anomaly.\cite{Okubo:1989vn}) Certainly, the case of zero dimensions is a
trivial realization of the general case of $8k$~dimensions, and therefore we
expect that in this case something similar to the global gauge anomaly
exists. Finally, if the integral is generalized to include several copies of
$\chi$ as $\int\prod_i\rmd\chi_i\,e^{\chi_i^TA\chi_i}$, the above argument
applies only when the number of~$\chi_i$ is odd.

Let us consider a possible application of the above observation. An integral
of the form~(\ref{one}) may emerge through the dimensional reduction to zero
dimensions\cite{Eguchi:1982nm} of higher-dimensional gauge theories which
contain the Majorana or Majorana-Weyl fermion. In this sense, this integral
represents a kind of matrix model. The simplest example that exhibits the
above phenomenon is obtained through the dimensional reduction of the
two-dimensional $\SO(2N)$ Yang-Mills theory with a single Majorana-Weyl
fermion in the fundamental representation. The dimensional reduction to zero
dimensions is given by
\begin{equation}
   \int\rmd\overline A\,\rmd A\,e^{\tr\{[A,\overline A]^2\}}\,
   \mathcal{O}(A,\overline A)\int\rmd\chi\,e^{\chi^TA\chi},
\label{five}
\end{equation}
where $A$ and $\overline A$ represent $2N\times 2N$ real anti-symmetric
traceless matrices. The function $\mathcal{O}(A,\overline A)$ is assumed to
consist of a combination of traces of a polynomial in $A$ and~$\overline A$.
Note that the Majorana-Weyl fermion in two dimensions has a single spinor
component, and it couples to only one of two components of the gauge
potential, due to the chiral projection. The measure
$\rmd\overline A\,\rmd A\,e^{\tr\{[A,\overline A]^2\}}$ and the
function $\mathcal{O}(A,\overline A)$ are invariant under the transformations
$A\to g^TAg$ and~$\overline A\to g^T\overline Ag$ for any
$g\in\mathop{\rm {}O}(2N)$. However, the pfaffian is not invariant, as
$\Pf(g^TAg)=-\Pf(A)$ for any $g$ satisfying $\det g=-1$, for example
$g=\mathop{\rm diag}(1,1,\ldots,1,-1)$. Thus the integral~(\ref{five})
identically vanishes. We may say that the system~(\ref{five}) suffers from
the global gauge anomaly in zero dimensions. The original two-dimensional
system, before the dimensional reduction, suffers from a local (or
perturbative) gauge anomaly, and thus is inconsistent. One might think that
nothing pathological happens after the dimensional reduction to zero
dimensions, but we have demonstrated that in fact this is not the case.

As a more complicated example, consider a zero-dimensional reduction of a
higher $d$-dimensional Yang-Mills theory which contains a single Majorana
fermion [thus $d$ is either 0, 1, 2, 3 or $4\pmod 8$]. We assume~$d\geq2$. A
dimensional reduction to zero dimensions is given by
\begin{equation}
   \int\left(\prod_\mu\rmd A_\mu\right)
   e^{S(A)}\,
   \mathcal{O}(A)\int\rmd\chi\,e^{\chi^TB\gamma_\mu A_\mu\chi},
\label{six}
\end{equation}
where $B$ is the charge conjugation matrix such that
$(B\gamma_\mu)^T=+B\gamma_\mu$, and $A_\mu$ denotes the gauge potential in
the gauge-group representation (which must be real and anti-symmetric) of the
Majorana fermion. As a whole, the combination~$B\gamma_\mu A_\mu$ is
anti-symmetric, as it should be.

If the gauge representation of the Majorana fermion is $r$~dimensional, the
Grassmann variable~$\chi$ is a $2^{[d/2]}r$-column vector. We assume the
structure
\begin{equation}
   g=h\otimes k,
\end{equation}
where the factor~$h$ acts on the spinor indices (and is thus a
$2^{[d/2]}\times2^{[d/2]}$ matrix) and $k$ acts on the gauge indices
(and is an $r\times r$ matrix). We thus have
\begin{equation}
   \det g=(\det h)^r\times(\det k)^{2^{[d/2]}}.
\end{equation}
The matrix~$g$ must be orthogonal. This requires that both $h$ and $k$ be
orthogonal, and thus $\det h=\pm1$ and $\det k=\pm1$. Since $2^{[d/2]}$
is always even for $d\geq2$, $\det g=-1$ can be realized only if $\det h=-1$
and $r$~is odd. Thus we have to find an appropriate~$h$ (which acts on the
spinor indices) such that $\det h=-1$ and the
measure~$(\prod_\mu\rmd A_\mu)\,e^{S(A)}$ is invariant under the action
of~$h$ on~$B\gamma_\mu A_\mu$. As a general feature of~$S(A)$, in addition to
the gauge invariance, we could assume invariance under $d$-dimensional
rotations and reflections. A rotation is induced by
$h=e^{\theta_{\mu\nu}[\gamma_\mu,\gamma_\nu]}$, and for this we have
$\det h=+1$. A reflection is induced by~$h\propto\gamma_\mu$, and it turns
out that $\det h=-1$ is possible only for $[d/2]=1$. Thus, cases in which we
can obtain $\det g=-1$ in this way are limited to~$d=2$ and $d=3$. For
example, consider a zero-dimensional reduction of the three-dimensional
$\SU(2N_c)$ Yang-Mills theory which couples to a single Majorana fermion in
the adjoint representation\footnote{This is a zero-dimensional reduction of
the $\mathcal{N}=1$ $\SU(2N_c)$ super Yang-Mills theory in three dimensions.
This system was studied~\cite{Aoki:1998vn} in the context of the IIB matrix
model\cite{Ishibashi:1996xs} for the superstring theory, and it is known
that the partition function vanishes (i)~to one-loop order for all $N_c$ and
(ii)~exactly for $N_c=1$. Our present argument provides a simple explanation
of this fact for all~$N_c$.} (i.e., the dimension of the gauge
representation~$r=4N_c^2-1$ is always odd). In the representation
$\gamma_0=\sigma_1$, $\gamma_1=\sigma_2$, $\gamma_2=\sigma_3$ and
$B=\sigma_2$, where $\sigma_i$ is a Pauli matrix, we can set
$g=\gamma_0\otimes 1$. Then we have $\det g=-1$, and
$B\gamma_\mu A_\mu\to g^TB\gamma_\mu A_\mu g$ induces the reflection
$A_0\to-A_0$, $A_1\to A_1$ and~$A_2\to A_2$. The ``action''
$S(A)=\tr\{[A_\mu,A_\nu][A_\mu,A_\nu]\}$ is invariant under this reflection.
This shows that the integral~(\ref{six}) with any {\it reflection
invariant\/} function $\mathcal{O}$ vanishes. In particular, the ``partition
function'', for which $\mathcal{O}=1$, identically vanishes. Thus, any
expectation value normalized by the partition function is either
indeterminate or infinite. In a similar way, we find the same situation for
the dimensional reduction of the two-dimensional $\SU(2N_c)$ Yang-Mills
theory that couples to a single Majorana fermion in the adjoint
representation.

In the present note, we have considered a zero-dimensional analogue of the
global gauge anomaly. We found that a zero-dimensional reduction of some
two- and three-dimensional gauge field theories suffers from this anomaly.
Certainly Eq.~(\ref{one}) does not constitute a field theory. However,
Eq.~(\ref{three}) is quite suggestive for a possible global gauge anomaly in
higher-dimensional gauge field theories that contain the Majorana fermion.
The functional integral over the Majorana fermion in such a system results
(at least formally) in a pfaffian $\Pf(B\gamma_\mu D_\mu)$, where $D_\mu$ is
the covariant derivative. In $8k$ and $8k+1$~dimensions, there is no obvious
way to regularize this pfaffian in a gauge invariant manner,\footnote{This
is also the case in lattice formulations of the Majorana
fermion\cite{Inagaki:2004ar}. This was one of original motivations of the
present work.} because a single Majorana fermion in $8k$ or
$8k+1$~dimensions cannot have a Lorentz invariant mass term. There is thus a
possibility that the Majorana fermion in such number of dimensions exhibits
the global gauge anomaly. Equation~(\ref{three}) indicates that this occurs
if there exists an element~$g$ of the gauge transformation satisfying
\begin{equation}
   \Det g=-1,
\end{equation}
where $\Det$ denotes the determinant with respect to spinor, group and
spacetime indices. The matrix~$g$ must be orthogonal (because the gauge
representation of the Majorana fermion must be real), and thus
$(\Det g)^2=\Det(g^Tg)=1$ and (at least formally) $\Det g=\pm1$. Therefore,
if we define $\Det g=+1$ for the identity~$g=1$, the pfaffian is invariant
under gauge transformations which can be smoothly deformed into the
identity. It is thus necessary that $\pi_d(G)$ be non-trivial for
$\Det g=-1$, as in the case of the global gauge anomaly associated with Weyl
fermions.\cite{Witten:fp} We believe that it is possible to evaluate the
determinant~$\Det g$ (with an appropriate regularization) and examine the
global gauge anomaly associated with the Majorana fermion in $8k$- and
$8k+1$-dimensions. We hope to come back to this problem in the near future.

\section*{Acknowledgements}
This work is basically a result of discussions at the Niigata
University-Yamagata University joint school (YITP-S-05-02). We would like to
thank the participants for stimulating discussions and the Yukawa Institute
for Theoretical Physics at Kyoto University for financial support of the
school. H.~S. would like to thank Tsukasa Tada for informative discussions.
This work is supported in part by the MEXT
Grants-in-Aid for Scientific Research
Nos.~13135203, 17540242 and~17043004.

\section*{Note added}
It is possible to show~\cite{Hayakawa:2006fd} that the Majorana fermion in
1~dimension in fact suffers from the global gauge anomaly, following a
calculation in exercise~5.8 of the first reference of
ref.~\citen{Jackiw:1983nv}. We would like to thank Roman Jackiw for calling
this reference to our attention.


\begin{thebibliography}{99}
  
\bibitem{Witten:fp}
E.~Witten,
\PLB{117,1982,324}.
\\
S.~Elitzur and V.~P.~Nair,
\NPB{243,1984,205}.
\\
See also,
E.~Witten,
\NPB{223,1983,422};
\NPB{223,1983,433}.

\bibitem{Holman:ef}
R.~Holman and T.~W.~Kephart,
\PLB{167,1986,417}.

\bibitem{Okubo:1989vn}
S.~Okubo and Y.~Tosa,
\PRD{40,1989,1925}.

\bibitem{Eguchi:1982nm}
T.~Eguchi and H.~Kawai,
\PRL{48,1982,1063}.

\bibitem{Aoki:1998vn}
H.~Aoki, S.~Iso, H.~Kawai, Y.~Kitazawa and T.~Tada,
\PTP{99,1998,713}; hep-th/9802085, Sec.~2 and Appendix~B.
\\
See also, T.~Suyama and A.~Tsuchiya,
\PTP{99,1998,321}; hep-th/9711073.

\bibitem{Ishibashi:1996xs}
N.~Ishibashi, H.~Kawai, Y.~Kitazawa and A.~Tsuchiya,
\NPB{498,1997,467}; hep-th/9612115.

\bibitem{Inagaki:2004ar}
T.~Inagaki and H.~Suzuki,
\JHEP{07,2004,038}; hep-lat/0406026.
\\
H.~Suzuki,
\PTP{112,2004,855}; hep-lat/0407010.

\bibitem{Hayakawa:2006fd}
M.~Hayakawa and H.~Suzuki,
hep-th/0601026.

\bibitem{Jackiw:1983nv}
R.~Jackiw,
``Topological investigations of quantized gauge theories,''
in S.~Treiman, R.~Jackiw, B.~Zumino and E.~Witten, Current Algebras and
Anomalies (Princeton University Press, Princeton, NJ/World Scientific,
Singapore, 1985). See also,
G.~V.~Dunne, R.~Jackiw and C.~A.~Trugenberger,
\PRD{41,1990,661}.

\end{thebibliography}
\end{document}